# Fractional Differential Forms II

Kathleen Cotrill-Shepherd and Mark Naber

Department of Mathematics

Monroe County Community College

Monroe, Michigan, 48161-9746

mnaber@monroeccc.edu

**ABSTRACT**

This work further develops the properties of fractional differential forms. In particular, finite dimensional subspaces of fractional form spaces are considered. An inner product, Hodge dual, and covariant derivative are defined. Coordinate transformation rules for integral order forms are also computed. Matrix order fractional calculus is used to define matrix order forms. This is achieved by combining matrix order derivatives with exterior derivatives. Coordinate transformation rules and covariant derivative for matrix order forms are also produced. The Poincaré lemma is shown to be true for exterior fractional differintegrals of all orders excluding those whose orders are non-diagonalizable matrices.

## 1. Introduction.

The use of differential forms in physics, differential geometry, and applied mathematics is well known and wide spread. There is also a growing use of differential forms in mathematical finance [23], [3]. Many partial differential equations are expressible in terms of differential forms. This notation allows for insights and a geometric understanding of the quantities involved. Recently Maxwell's equations have been generalized using fractional derivatives, in part, to better understand multipole moments [5]-[8]. Kobelev has also generalized Maxwell's and Einstein's equations with fractional derivatives for study on multifractal sets [10]-[12]. Many of the differential equations of finance have also been fractionalized to better predict the pricing of options and equities in financial markets [1], [22], [24], [25]. Fractional calculus is also being applied to statistics [18]-[20]. Hence, there is a need to further develop the notion of a fractional differential form to better understand fractional differential equations, fractional line elements, and fractal geometry.

In an earlier work [4] (hereafter referred to as FDF I) the study of fractional differential forms was initiated. The purpose of that study was to combine Riemann-Liouville fractional calculus with exterior calculus on an n-dimensional Euclidean space. FDF I focused primarily on differential order fractional forms. In this paper the properties of differential order fractional forms are extended. Integral and matrix order forms are also introduced and studied. Following Oldham and Spanier [17] the term fractional differintegral form shall be used when the result applies to both differential and integral order forms.

In section 2 some algebraic properties are worked out for various finite dimensional subspaces of the $F(\nu,m,n)$, vector spaces defined in FDF I. In particular, a Euclidean inner product and Hodge dual are constructed for these finite dimensional subspaces and then for the infinite dimensional $F(\nu,m,n)$. In section 3 some differential properties are worked out for the finite dimensional subspaces of $\mathrm{F}(\nu,\mathrm{m},\mathrm{n})$. The fractional Poincaré lemma is considered for all orders of the fractional exterior differintegral, including orders that represent fractional integration. Coordinate transformation rules for integral order forms are constructed. An example of a change of coordinates for integral order forms is also provided. In section 4, matrix order differintegrals are used to define matrix order forms (see the appendix for the definition of matrix order derivatives and integrals). Coordinate transformation rules and a covariant derivative are also constructed. The fractional Poincaré lemma is extended to matrix order forms and found to be true provided that the matrix representing the order is diagonalizable.

As in FDF I the coordinate index will be a sub-script rather than the traditional super-script. The summation convention will not be used so as to maintain some clarity with the myriad of indices to be encountered. Throughout this paper $\{x_i\}$ shall represent Cartesian coordinates and, $\{y_i\}$ and $\{z_i\}$ shall represent arbitrary curvilinear coordinates on an n-dimensional Euclidean space, $\mathrm{E}^n$. $a_i$ shall denote the initial point for evaluation of the fractional differintegrals in Cartesian coordinates and, $\tilde{a}_i$ and $\tilde{\tilde{a}}_i$ in the curvilinear coordinates.

Before beginning it is necessary to correct an error and two typographical errors from FDF I that may cause confusion. In FDF I equation (25) is in error and equations (26) and (27) have miss typed indices. The equations should read, respectively,

$$\left\{dx_i^{\mu_1} \wedge dx_j^{\mu_2} \;\middle|\; i,j \in \{1,\ldots,n\}, \forall \mu_1,\mu_2 \geq 0, \ni \mu_1 + \mu_2 = \nu\right\}, \qquad (1)$$

$$\beta = \sum_{i=1}^{n} \sum_{j=1}^{n} \int_0^{\nu} \left(\beta_{ij}(\nu_i,\nu-\nu_i)\, dx_i^{\nu_i} \wedge dx_j^{\nu-\nu_i}\right) d\nu_i, \qquad (2)$$

$$\beta = \sum_{i=1}^{n}\sum_{j=1}^{n}\sum_{k=1}^{n}\int_0^{\nu}\int_0^{\nu-\nu_i}\left(\beta_{ijk}(\nu_i,\nu-\nu_j,\nu-\nu_i-\nu_j)dx_i^{\nu_i}\wedge dx_j^{\nu_j}\wedge dx_k^{\nu-\nu_i-\nu_j}\right)d\nu_j d\nu_i . \quad (3)$$

A more compact notation for fractional differintegrals is also adopted. The notation for fractional integrals and derivatives is given below.

$${}_{a_i}\mathrm{D}_x^{-\lambda}\mathrm{f}(x) = \frac{1}{\Gamma(\lambda)}\int_{a_i}^{x}\frac{\mathrm{f}(\xi)\mathrm{d}\xi}{(x-\xi)^{1-\lambda}} \;, \quad \mathrm{Re}(\lambda) > 0 \quad (4)$$

$${}_{a_i}\mathrm{D}_x^{\lambda}\mathrm{f}(x) = \frac{\partial^n}{\partial x^n}\left[\frac{1}{\Gamma(n-\lambda)}\int_{a_i}^{x}\frac{\mathrm{f}(\xi)\mathrm{d}\xi}{(x-\xi)^{\lambda-n+1}}\right] \qquad \frac{\mathrm{Re}(\lambda) \geq 0}{\mathrm{n} > \mathrm{Re}(\lambda)\ (\mathrm{n\ is\ whole})} \quad (5)$$

The parameter λ is the order of the integral or derivative and is allowed to be complex. In the appendix it is shown that λ may assume matrix order values as well. Equation (4) is a fractional integral and equation (5) is a fractional derivative. Taken together, the operation is referred to as differintegration (see page 61 of [17]).

Equations (6) – (13) are the basic identities from fractional calculus. In the following, m is a whole number and p and q are complex numbers whose real part is greater than zero.

$$\frac{\partial^m}{\partial x_i^{\,m}}\,{}_{a_i}\mathrm{D}_{x_i}^{q}\,\mathrm{f}(x) = {}_{a_i}\mathrm{D}_{x_i}^{q+m}\mathrm{f}(x) \quad (6)$$

$${}_{a_i}\mathrm{D}_{x_i}^{q}\,{}_{a_i}\mathrm{D}_{x_i}^{-q}\mathrm{f}(x) = \mathrm{f}(x) \quad (7)$$

$$ {}_{a_i}\mathrm{D}_{x_i}^{-q}\,{}_{a_i}\mathrm{D}_{x_i}^{q}\,\mathrm{f}(x) \neq \mathrm{f}(x) \tag{8} $$

$$ {}_{a_i}\mathrm{D}_{x_i}^{p}\,{}_{a_i}\mathrm{D}_{x_i}^{-q}\mathrm{f}(x) = {}_{a_i}\mathrm{D}_{x_i}^{p-q}\mathrm{f}(x) \tag{9} $$

$$ {}_{a_i}\mathrm{D}_{x_i}^{p}\,{}_{a_i}\mathrm{D}_{x_i}^{q}\,\mathrm{f}(x) = {}_{a_i}\mathrm{D}_{x_i}^{p+q}\mathrm{f}(x) - \sum_{j=1}^{k} {}_{a_i}\mathrm{D}_{x_i}^{q-j}\mathrm{f}(x)\Big|_{x_i=a_i} \frac{(x_i-a_i)^{-p-j}}{\Gamma(1-p-j)} \tag{10} $$

$$ {}_{a_i}\mathrm{D}_{x_i}^{-p}\,{}_{a_i}\mathrm{D}_{x_i}^{q}\,\mathrm{f}(x) = {}_{a_i}\mathrm{D}_{x_i}^{q-p}\mathrm{f}(x) - \sum_{j=1}^{k} {}_{a_i}\mathrm{D}_{x_i}^{q-j}\mathrm{f}(x)\Big|_{x_i=a_i} \frac{(x_i-a_i)^{p-j}}{\Gamma(1+p-j)} \tag{11} $$

$$ {}_{a_i}\mathrm{D}_{x_i}^{-p}\left({}_{a_i}\mathrm{D}_{x_i}^{-q}(\mathrm{f}(x))\right) = {}_{a_i}\mathrm{D}_{x_i}^{-(p+q)}(\mathrm{f}(x)) \tag{12} $$

$$ {}_{a_i}\mathrm{D}_{x_i}^{q}\,(\mathrm{f}(x)\mathrm{g}(x)) = \sum_{s=0}^{\infty} \binom{q}{s} \left({}_{a_i}\mathrm{D}_{x_i}^{q-s}\mathrm{f}(x)\right)\left(\partial_{x_i}^{s}\,\mathrm{g}(x)\right) \tag{13} $$

In equations (10) and (11) k is the first whole number $\geq \mathrm{Re}(q)$. The above formulae and definitions can be found in [16], [17], [21].

## 2. Algebraic Properties.

In this section the restriction that the order of the fractional differintegrals be real is made. Many of the results will clearly be valid for complex order differintegrals as well. To begin, define a vector space $\mathrm{G}(\nu,\mathrm{n})$ that is a finite dimensional subspace of $\mathrm{F}(\nu,\mathrm{m},\mathrm{n})$ (this is defined in FDF I) such that,

$$\mathrm{G}(\nu_1,\nu_2,\ldots,\nu_m,\mathrm{n}) = \prod_{i=1}^{m} \mathrm{G}(\nu_i,\mathrm{n}), \tag{14}$$

and,

$$\mathrm{F}(\nu,\mathrm{m},\mathrm{n}) = \bigcup_{\sum_{i=1}^{m} \nu_i = \nu} \mathrm{G}(\nu_1,\nu_2,\ldots,\nu_m,\mathrm{n}). \tag{15}$$

If $\nu \geq 0$ then $\nu_i \geq 0$ for all $i$ is imposed. If $\nu < 0$ then $\nu_i < 0$ for all $i$ is imposed.

$$\alpha \in \mathrm{G}(\nu_1,\nu_2,\ldots,\nu_m,\mathrm{n}) \Rightarrow \alpha = \sum_{i_1 \cdots i_m = 1}^{n} \alpha_{i_1 \cdots i_m} \mathrm{d}x_{i_1}^{\nu_1} \wedge \cdots \wedge \mathrm{d}x_{i_m}^{\nu_m}. \tag{16}$$

Some of the terms in the sum may be zero if any of the orders of the coordinate differentials are the same. For example, $\mathrm{d}x_i^{1/2} \wedge \mathrm{d}x_i^{1/2} = 0$. The dimensions of these vector spaces are

$$\dim(\mathrm{G}(\nu_1,\mathrm{n})) = \mathrm{n}, \tag{17}$$

$$\dim(\mathrm{G}(\nu_1,\nu_1,\mathrm{n})) = \begin{pmatrix} n \\ 2 \end{pmatrix}, \tag{18}$$

$$\dim(\mathrm{G}(\nu_1,\nu_1,\ldots,\nu_1,\mathrm{n})) = \begin{pmatrix} n \\ m \end{pmatrix} \quad m \leq n. \tag{19}$$

In equation (19) there are m coordinate differentials of order $\nu_1$.

$$\dim(\mathrm{G}(\nu_1,\nu_2,\mathrm{n})) = n^2 \qquad \nu_1 \neq \nu_2 \tag{20}$$

$$\dim(\mathrm{G}(\nu_1,\nu_2,\ldots,\nu_m,\mathrm{n})) = n^m \tag{21}$$

$$\dim(\mathrm{G}(\nu_1,\ldots\,\nu_1,\nu_2,\ldots,\nu_2,\ldots,\nu_m,\ldots\,\nu_m,\mathrm{n})) = \binom{n}{p_1}\binom{n}{p_2}\ldots\binom{n}{p_m} \tag{22}$$

In equations (21) and (22) the $\nu_i$ are distinct for $i \in \{1,2,\ldots,m\}$. In equation (22) there are $p_1$ coordinate differentials of order $\nu_1$, $p_2$ coordinate differentials of order $\nu_2$, etc., i.e.

$$\mathrm{G}(\nu_1,\ldots\,\nu_1,\nu_2,\ldots,\nu_2,\ldots,\nu_m,\ldots\,\nu_m,\mathrm{n}) = \prod_{j=1}^{m}\left(\prod_{i=1}^{p_j}\mathrm{G}(\nu_i,\mathrm{n})\right). \tag{23}$$

An inner product on $\mathrm{G}(\nu_1,\mathrm{n})$ can be defined in the usual way (see [2], [9], and [14]) and then built up to an inner product on the higher dimensional subspaces. Let $\alpha,\ \beta \in \mathrm{G}(\nu_1,\mathrm{n})$ with $\alpha = \sum_{\mathrm{i}=1}^{\mathrm{n}} \alpha_i \mathrm{d}x_i^{\nu_1}$ and $\beta = \sum_{\mathrm{i}=1}^{\mathrm{n}} \beta_i \mathrm{d}x_i^{\nu_1}$. The Euclidean inner product on $\mathrm{E}^n$ for $\mathrm{G}(\nu_1,\mathrm{n})$ is

$$(\alpha,\beta) = \sum_{i=1}^{n} \alpha_i\, \beta_i\,. \tag{24}$$

In curvilinear coordinates equation (24) becomes

$$(\alpha,\beta) = \sum_{i,j=1}^{n} \alpha_i\, \beta_j\, g^{ij}(y,\nu_1)\,. \tag{25}$$

The metric $g^{ij}(y,\nu_1)$ is defined in equation (67) of FDF I. The $\mathrm{d}x_i^{\nu_1}$ form an orthonormal basis on $\mathrm{G}(\nu_1,\mathrm{n})$.

$$(\mathrm{d}x_i^{v_1},\ \mathrm{d}x_j^{v_1}) = \delta_{ij} \tag{26}$$

If it is assumed that the $v_i$ are distinct for $i \in \{1, \ldots, m\}$, then the inner product on $\mathrm{G}(v_1, v_2, \ldots, v_m, \mathrm{n})$ is given as follows. Let $\alpha = \sum_{i_1 \cdots i_m = 1}^{n} \alpha_{i_1 \cdots i_m} \mathrm{d}x_{i_1}^{v_1} \wedge \cdots \wedge \mathrm{d}x_{i_m}^{v_m}$ and $\beta = \sum_{i_1 \cdots i_m = 1}^{n} \beta_{i_1 \cdots i_m} \mathrm{d}x_{i_1}^{v_1} \wedge \cdots \wedge \mathrm{d}x_{i_m}^{v_m}$ then,

$$(\alpha, \beta) = \sum_{i_1 \cdots i_m = 1}^{n} \alpha_{i_1 \cdots i_m} \beta_{i_1 \cdots i_m}. \tag{27}$$

In curvilinear coordinates equation (27) becomes

$$(\alpha, \beta) = \sum_{i_1 \cdots i_m = 1}^{n} \sum_{j_1 \cdots j_m = 1}^{n} \alpha_{i_1 \cdots i_m} \beta_{j_1 \cdots j_m} g^{i_1 j_1}(y, v_1) \cdots g^{i_m j_m}(y, v_m) \tag{28}$$

An orthonormal basis for $\mathrm{G}(v_1, v_2, \ldots, v_m, \mathrm{n})$ is given by,

$\left\{ \mathrm{d}x_{i_1}^{v_1} \wedge \cdots \wedge \mathrm{d}x_{i_m}^{v_m} \mid i_1, i_2, \ldots i_m \in (1, 2, \ldots, n) \right\}$ and,

$$(\mathrm{d}x_{i_1}^{v_1} \wedge \cdots \wedge \mathrm{d}x_{i_m}^{v_m}, \mathrm{d}x_{j_1}^{v_1} \wedge \cdots \wedge \mathrm{d}x_{j_m}^{v_m}) = \delta_{i_1 j_1} \cdots \delta_{i_m j_m}. \tag{29}$$

By extension, an orthonormal basis for $\mathrm{G}(v_1, \ldots v_1, v_2, \ldots, v_2, \ldots, v_m, \ldots v_m, \mathrm{n})$, where there are $p_i$ differentials of order $v_i$, and the $v_i$ are distinct for $i \in \{1,\ 2,\ \ldots,\ m\}$, is given by,

$\left\{ \mathrm{d}x_{{}_1 i_1}^{v_1} \wedge \cdots \wedge \mathrm{d}x_{{}_1 i_{p_1}}^{v_1} \wedge \mathrm{d}x_{{}_2 i_1}^{v_2} \wedge \cdots \wedge \mathrm{d}x_{{}_2 i_{p_2}}^{v_2} \wedge \cdots \wedge \mathrm{d}x_{{}_m i_1}^{v_m} \wedge \cdots \wedge \mathrm{d}x_{{}_m i_{p_m}}^{v_m} \mid {}_1 i_1, {}_1 i_2, \ldots, {}_m i_{p_m} \in (1, 2, \ldots, n) \right\}.$

The coordinate indices assume all possible combinations with the constraint that

$1 \le {}_1i_1 < {}_1i_2 < \cdots < {}_1i_{p_1} \le n$, $1 \le {}_2i_1 < {}_2i_2 < \cdots < {}_2i_{p_2} \le n$, . . . , $1 \le {}_mi_1 < {}_mi_2 < \cdots < {}_mi_{p_m} \le n$.

Let $\alpha \in G(\nu_1,\ldots,\nu_m,n) = \prod_{j=1}^{m}\left(\prod_{i=1}^{p_j} G(\nu_i,n)\right)$ and $\beta \in G(\mu_1,\ldots,\mu_l,n) = \prod_{j=1}^{l}\left(\prod_{i=1}^{q_j} G(\mu_i,n)\right)$.

Then the exterior product of $\alpha$ and $\beta$ obeys

$$\alpha \wedge \beta = (-1)^{\left(\sum_{i=1}^{m} p_i\right)\left(\sum_{j=1}^{l} q_j\right)} \beta \wedge \alpha. \tag{30}$$

The Hodge dual on $\mathrm{G}(\nu_1,\nu_1,\ldots,\nu_1,\mathrm{n})$, in Cartesian coordinates, is constructed in the same way as in ordinary exterior algebra (see e.g., page 287 of [14] and page 108 of [21]). Specify a basis with an orientation $\{\mathrm{d}x_1^{\nu_1}, \mathrm{d}x_2^{\nu_1}, \ldots, \mathrm{d}x_n^{\nu_1}\}$ for $\mathrm{G}(\nu_1,\mathrm{n})$. The $\mathrm{d}x_i^{\nu_1}$ form an orthonormal basis on $\mathrm{G}(\nu_1,\mathrm{n})$, $(\mathrm{d}x_i^{\nu_1}, \mathrm{d}x_j^{\nu_1}) = \delta_{ij}$. $*$ denotes the Hodge dual,

$$*(\mathrm{d}x_{i_1}^{\nu_1} \wedge \cdots \wedge \mathrm{d}x_{i_p}^{\nu_1}) = \mathrm{d}x_{j_1}^{\nu_1} \wedge \cdots \wedge \mathrm{d}x_{j_{n-p}}^{\nu_1}, \tag{31}$$

where $(i_1, \ldots, i_p, j_1, \ldots, j_{n-p})$ is an even permutation of $(1, \ldots, n)$. In curvilinear coordinates the Hodge dual of $\alpha = \sum_{i_1 \cdots i_p=1}^{n} \alpha_{i_1 \cdots i_p} dy_{i_1}^{\nu_1} \wedge \cdots \wedge dy_{i_p}^{\nu_1}$ is also constructed just as it is in exterior calculus but respecting the fractional coordinate transformation rules (see equation (57) of FDF I). Let $\mathrm{J}(\nu) = \det(\mathrm{J}^{\,i}_{\;j}(x,y,\nu))$ and $1/\mathrm{J}(\nu) = \det(\mathrm{J}^{\,j}_{\;i}(y,x,\nu))$. Let $\varepsilon^{j_1 \cdots j_n}$ denote the Levi Civita permutation symbol, i.e. $\varepsilon^{j_1 \cdots j_n}\varepsilon_{j_1 \cdots j_n} = n!$. Then, in index notation,

$$ {}^*\alpha = \frac{1}{(n-p)!}\sum_{j_1\cdots j_p=1}^{n}\frac{\varepsilon^{j_1\cdots j_n}}{\mathrm{J}(v)}\alpha_{j_1\cdots j_p}, \tag{32} $$

$$ {}^{**}\alpha = (-1)^{p(n-p)}\alpha. \tag{33} $$

If desired, $\sqrt{g(v)} = \sqrt{\det(g_{ij}(x,y,v))}$ may be used in place of $\mathrm{J}(v) = \det(\mathrm{J}^{i}_{j}(x,y,v))$.

A dual for $\mathrm{G}(v_1,\ldots v_1,v_2,\ldots,v_2,\ldots,v_m,\ldots v_m,\mathrm{n})$ can be constructed by noting that,

$$ \mathrm{G}(v_1,\ldots v_1,v_2,\ldots,v_2,\ldots,v_m,\ldots v_m,\mathrm{n}) = \prod_{j=1}^{m}\left(\prod_{i=1}^{p_j}\mathrm{G}(v_i,\mathrm{n})\right). \tag{34} $$

Then,

$$ {}^*\mathrm{G}(v_1,\ldots,v_1,v_2,\ldots,v_2,\ldots,v_m,\ldots v_m,\mathrm{n}) = \prod_{j=1}^{m}{}^*\left(\prod_{i=1}^{p_j}\mathrm{G}(v_i,\mathrm{n})\right), \tag{35} $$

$$ {}^*\mathrm{G}(v_1,\ldots,v_1,v_2,\ldots,v_2,\ldots,v_m,\ldots,v_m,\mathrm{n}) = \prod_{j=1}^{m}\left(\prod_{i=1}^{n-p_j}\mathrm{G}(v_i,\mathrm{n})\right). \tag{36} $$

For example consider,

$$ \alpha = \sum_{{}_1i_1\cdots {}_1i_{p_1}=1}^{n}\cdots\sum_{{}_mi_1\cdots {}_mi_{p_m}=1}^{n}\alpha_{{}_1i_1\cdots {}_1i_{p_1}\cdots {}_mi_1\cdots {}_mi_{p_m}}\, dy^{v_1}_{{}_1i_1}\wedge\cdots\wedge dy^{v_1}_{{}_1i_{p_1}}\wedge\cdots\wedge dy^{v_m}_{{}_mi_1}\wedge\cdots\wedge dy^{v_m}_{{}_mi_{p_m}}. \tag{37} $$

In index notation,

$$ {}^{*}\alpha = \sum_{{}_1 i_1 \cdots {}_1 i_{p_1}=1}^{n} \cdots \sum_{{}_m i_1 \cdots {}_m i_{p_m}=1}^{n} \left( \prod_{k=1}^{m} \frac{1}{(n-p_k)!} \frac{\varepsilon^{{}_k i_1 \cdots {}_k i_n}}{\mathrm{J}(\nu_k)} \right) \alpha_{{}_1 i_1 \cdots {}_1 i_{p_1} \cdots {}_m i_1 \cdots {}_m i_{p_m}} , \tag{38} $$

$$ {}^{**}\alpha = \left( \prod_{k=1}^{m} (-1)^{p_k (n-p_k)} \right) \alpha . \tag{39} $$

For the discussion of the inner product and Hodge dual for $\mathrm{F}(\nu,\mathrm{m},\mathrm{n})$ the restriction that $\nu \geq 0$ is made. The inner product and Hodge dual for $\mathrm{F}(\nu,1,\mathrm{n})$ would be the same as for $G(\nu,n)$. For $\mathrm{F}(\nu,2,\mathrm{n})$, the basis elements are made up of two coordinate differentials.

$$ \left\{ dx_i^{\mu_1} \wedge dx_j^{\mu_2} \;\middle|\; i,j \in \{1,\ldots,n\}, \forall \mu_1,\mu_2 \geq 0, \ni \mu_1 + \mu_2 = \nu \right\} \tag{40} $$

Consider two arbitrary elements of $\mathrm{F}(\nu,2,\mathrm{n})$

$$ \alpha = \sum_{i,j=1}^{n} \int_0^{\nu} \left( \alpha_{ij}(\nu_i, \nu - \nu_i) dx_i^{\nu_i} \wedge dx_j^{\nu-\nu_i} \right) d\nu_i , \tag{41} $$

$$ \beta = \sum_{i,j=1}^{n} \int_0^{\nu} \left( \beta_{ij}(\nu_i, \nu - \nu_i) dx_i^{\nu_i} \wedge dx_j^{\nu-\nu_i} \right) d\nu_i . \tag{42} $$

The inner product on $\mathrm{F}(\nu,2,\mathrm{n})$ is,

$$ (\alpha,\beta) = \sum_{i,j=1}^{n} \int_0^{\nu} \left( \alpha_{ij}(\nu_i, \nu - \nu_i) \beta_{ij}(\nu_i, \nu - \nu_i) \right) d\nu_i . \tag{43} $$

In curvilinear coordinates equation (43) becomes,

$$(\alpha,\beta) = \sum_{i,j=1}^{n} \int_0^{\nu} \big(\alpha_{i\,k}(\nu_i,\nu-\nu_i)\beta_{j\,l}(\nu_i,\nu-\nu_i)\big) g^{ij}(y,\nu_i) g^{kl}(y,\nu-\nu_i)\; d\nu_i \,. \tag{44}$$

An inner product on $\mathrm{F}(\nu,\mathrm{m},\mathrm{n})$ would involve m-1 integrals and m summations.

The problem with doing a Hodge dual on $\mathrm{F}(\nu,2,\mathrm{n})$ is that there is a subspace where the orders of the coordinate differentials are the same. That is $\mathrm{G}(\nu/2,\nu/2,\mathrm{n})$. On this subspace the Hodge dual mapping will map into objects with (n – 2) coordinate differentials. On the rest of $\mathrm{F}(\nu,2,\mathrm{n})$ the Hodge dual mapping will map into objects with (2n – 2) coordinate differentials. To avoid this problem consider an arbitrary element of $\mathrm{F}(\nu,2,\mathrm{n})$ that has no component in $\mathrm{G}(\nu/2,\nu/2,\mathrm{n})$.

$$\begin{aligned}\alpha \;=\; & \lim_{\varepsilon\to\frac{\nu}{2}^-} \sum_{i,j=1}^{n} \int_0^{\varepsilon} \big(\alpha_{i\,j}(\nu_i,\nu-\nu_i) dy_i^{\nu_i} \wedge dy_j^{\nu-\nu_i}\big)\, d\nu_i \\ & + \lim_{\varepsilon\to\frac{\nu}{2}^+} \sum_{i,j=1}^{n} \int_{\varepsilon}^{\nu} \big(\alpha_{i\,j}(\nu_i,\nu-\nu_i) dy_i^{\nu_i} \wedge dy_j^{\nu-\nu_i}\big)\, d\nu_i\end{aligned} \tag{45}$$

Then, in index notation, the Hodge dual of $\alpha$ is,

$$\begin{aligned}{}^*\alpha \;=\; & \lim_{\varepsilon\to\frac{\nu}{2}^-} \int_0^{\varepsilon} \left( \sum_{i_1,j_1=1}^{n} \left(\frac{1}{(n-1)!}\right)^2 \frac{\varepsilon^{i_1 \cdots i_n}}{\mathrm{J}(\nu_{i_1})} \frac{\varepsilon^{j_1 \cdots j_n}}{\mathrm{J}(\nu-\nu_{i_1})} \alpha_{i_1 j_1}(\nu_{i_1},\nu-\nu_{i_1}) \right) d\nu_{i_1} \\ & + \lim_{\varepsilon\to\frac{\nu}{2}^+} \int_{\varepsilon}^{\nu} \left( \sum_{i_1,j_1=1}^{n} \left(\frac{1}{(n-1)!}\right)^2 \frac{\varepsilon^{i_1 \cdots i_n}}{\mathrm{J}(\nu_{i_1})} \frac{\varepsilon^{j_1 \cdots j_n}}{\mathrm{J}(\nu-\nu_{i_1})} \alpha_{i_1 j_1}(\nu_{i_1},\nu-\nu_{i_1}) \right) d\nu_{i_1}\end{aligned} \tag{46}$$

$$^{**}\alpha = (-1)^{2(n-2)}\alpha = \alpha \tag{47}$$

The Hodge dual for $\mathrm{F}(\nu,\mathrm{m},\mathrm{n})$ is constructed in the same way, with the restriction of removing subspaces where the orders of the coordinate differentials are the same. Note that $m \le n$ is also required.

**3. Differential Properties.**

To begin this section, it is desired to determine the values of $\nu$ and $\alpha$ for which the fractional Poincaré identity is always true. $\nu$ being the order of the exterior fractional differintegral and $\alpha$ being the object that is acted upon.

$$\mathrm{d}^{\nu}\mathrm{d}^{\nu}(\alpha) = 0 \tag{48}$$

Let $\alpha \in \mathrm{G}(\nu_1,\ldots,\nu_1,\nu_2,\ldots,\nu_2,\ldots,\nu_m,\ldots,\nu_m,\mathrm{n})$ and denote the elements of an orthonormal basis by $\sigma_l$. To be specific, $\alpha = \sum_{l=1}^{N}\alpha_l\sigma_l$, $N = \binom{n}{p_1}\binom{n}{p_2}\cdots\binom{n}{p_m}$,

$\sigma_l \in \left\{\mathrm{d}x^{\nu_1}_{1\,i_1} \wedge \cdots \wedge \mathrm{d}x^{\nu_1}_{1\,i_{p_1}} \wedge \cdots \wedge \mathrm{d}x^{\nu_m}_{m\,i_1} \wedge \cdots \wedge \mathrm{d}x^{\nu_m}_{m\,i_{p_m}} \mid {}_1i_1, {}_1i_2, \ldots {}_m i_{p_m} \in (1,2,\ldots,n)\right\}$, and the $p_1 \cdots p_m$ are defined in equations (22) and (23). Substituting this into equation (48) and expanding the fractional exterior differintegral gives,

$$\mathrm{d}^{\nu}\mathrm{d}^{\nu}(\alpha) = \sum_{l=1}^{N}\mathrm{d}^{\nu}\sum_{j=1}^{n}\mathrm{d}x_j^{\nu}\wedge {}_{a_j}\mathrm{D}^{\nu}_{x_j}(\alpha_l\sigma_l). \tag{49}$$

The product rule for fractional differintegrals is used to evaluate ${}_{a_j}\mathrm{D}^{\nu}_{x_j}(\alpha_l\sigma_l)$ (equation (13) of section 1).

$$ {}_{a_j}\mathrm{D}^{\nu}_{x_j}\left(\alpha_l\sigma_l\right) = \sum_{k=0}^{\infty}\binom{\nu}{k}\left({}_{a_j}\mathrm{D}^{\nu-k}_{x_j}\alpha_l\right)\left(\partial^{k}_{x_j}\sigma_l\right) \tag{50} $$

The second factor on the right-hand side will be zero for all k values except k = 0. Thus,

$$ {}_{a_j}\mathrm{D}^{\nu}_{x_j}\left(\alpha_l\sigma_l\right) = \left({}_{a_j}\mathrm{D}^{\nu}_{x_j}\alpha_l\right)\sigma_l, \tag{51} $$

and equation (49) becomes,

$$ \mathrm{d}^{\nu}\mathrm{d}^{\nu}(\alpha) = \sum_{l=1}^{N}\mathrm{d}^{\nu}\sum_{j=1}^{n}\left({}_{a_j}\mathrm{D}^{\nu}_{x_j}\alpha_l\right)\mathrm{d}x_j^{\nu}\wedge\sigma_l\,. \tag{52} $$

The second fractional exterior differintegral is expanded and reduced in the same manner yielding

$$ \mathrm{d}^{\nu}\mathrm{d}^{\nu}(\alpha) = \sum_{l=1}^{N}\sum_{j,k=1}^{n}\left({}_{a_k}\mathrm{D}^{\nu}_{x_k}\left({}_{a_j}\mathrm{D}^{\nu}_{x_j}\alpha_l\right)\right)\left(\mathrm{d}x_k^{\nu}\wedge\mathrm{d}x_j^{\nu}\wedge\sigma_l\right). \tag{53} $$

The factor $\mathrm{d}x_k^{\nu}\wedge\mathrm{d}x_j^{\nu}\wedge\sigma_l$ is antisymmetric in the k and j indices. The fractional differintegral operators are linear and with respect to different coordinates and are thus symmetric on the k and j indices. Hence, each term in the above summation is zero. Thus the fractional Poincaré lemma is true for all values of $\nu\in C$ and for all $\alpha\in \mathrm{G}(\nu_1,\ldots\nu_1,\nu_2,\ldots,\nu_2,\ldots,\nu_m,\ldots\nu_m,\mathrm{n})$ (assuming that the components of $\alpha$ are differintegrable to the appropriate orders). Recall that, $\mathrm{F}(\nu,\mathrm{m},\mathrm{n})$ is the union of all the $\mathrm{G}(\nu_1,\nu_2,\ldots,\nu_m,\mathrm{n})$ i.e.

$$\mathrm{F}(\nu,\mathrm{m,n}) = \bigcup_{\sum_{i=1}^{m} \nu_i = \nu} \mathrm{G}(\nu_1,\nu_2, \ldots ,\nu_m,\mathrm{n}). \tag{54}$$

Hence,

$$\mathrm{d}^\nu \mathrm{d}^\nu(\alpha) = 0 \quad \forall\ \nu \in C \text{ and } \forall \alpha \in \mathrm{F}(\nu,\mathrm{m,n}). \tag{55}$$

In FDF I the coordinate transformation rules were worked out for fractional differential forms. For fractional integral forms the coordinate transformation rule is less compact. The reason for this is that fractional derivatives ($\mathrm{Re}(\nu) \geq 0$) have a non-trivial kernel (see equation (31) of FDF I) while fractional integrals ($\mathrm{Re}(\nu) < 0$) do not. To begin, assume that the Cartesian coordinates, $\{x_i\}$, can be written smoothly in terms of the curvilinear coordinates, $\{y_i\}$, and note the expression for the exterior fractional differintegral in the two coordinate systems.

$$x_i = x_i(y) \tag{56}$$

$$\mathrm{d}^\nu = \sum_{i=1}^{n} \mathrm{d}x_i^\nu \, {}_{a_i}\mathrm{D}_{x_i}^\nu \tag{57}$$

$$\mathrm{d}^\nu = \sum_{j=1}^{n} \mathrm{d}y_j^\nu \, {}_{\tilde{a}_j}\mathrm{D}_{y_j}^\nu \tag{58}$$

The goal is to find a coordinate transformation matrix, $\mathrm{J}^i_j(x,y,\nu)$, that will express the $\mathrm{d}x_i^\nu$ in terms of the $\mathrm{d}y_i^\nu$.

$$\mathrm{d}x_i^{v} \;=\; \sum_{j=1}^{n}\mathrm{d}y_j^{v}\,\mathrm{J}_j^{i}(x,y,v) \tag{59}$$

To construct the matrix for the integral order forms let $k \in \{1,\ldots,n\}$. Then, the following will generate n linear equations for the $\mathrm{d}x_i^{v}$.

$$\sum_{i=1}^{n}\mathrm{d}x_i^{v}\;{}_{a_i}\mathrm{D}_{x_i}^{v}\left(x_k\right) \;=\; \sum_{j=1}^{n}\mathrm{d}y_j^{v}\;{}_{\tilde{a}_j}\mathrm{D}_{y_j}^{v}\left(x_k(y)\right) \tag{60}$$

Cramer's rule can now be used to find the matrix for the coordinate transformation. Define the $\mathrm{n}\times\mathrm{n}$ matrix A to be,

$$\mathrm{A}_{i\,k} \;=\; {}_{a_i}\mathrm{D}_{x_i}^{v}\left(x_k\right), \tag{61}$$

and a $1\times\mathrm{n}$ column vector,

$$\mathrm{b}_k \;=\; \sum_{j=1}^{n}\mathrm{d}y_j^{v}\;{}_{\tilde{a}_j}\mathrm{D}_{y_j}^{v}\left(x_k(y)\right). \tag{62}$$

Equation (60) can now be written as,

$$\sum_{i=1}^{n}\mathrm{d}x_i^{v}\,\mathrm{A}_{i\,k} = \mathrm{b}_\mathrm{k}. \tag{63}$$

The $\mathrm{d}x_i^{v}$ can now be solved for in terms of the $\mathrm{d}y_j^{v}$. Define ${}_i\mathrm{A}$ to be the matrix A with column number i replaced with the column vector $\mathrm{b}_k$. Then, via Cramer's rule,

$$\mathrm{d}x_i^{\nu} \;=\; \frac{\det\left({}_i\mathrm{A}\right)}{\det\left(\mathrm{A}\right)} \tag{64}$$

Equation (64) defines the coordinate transformation matrix for fractional integral forms.

$$\mathrm{d}x_i^{\nu} \;=\; \sum_{j=1}^{n} \mathrm{d}y_j^{\nu} J^i_j(x,y,\nu) \tag{65}$$

As an example, consider the transformation from Cartesian to polar coordinates with the order $\nu = -1$ and the initial point for the fractional integrals is the origin.

$$\mathrm{d}x^{-1} = \frac{2\tan(\theta)-1}{3}\mathrm{d}r^{-1} - \frac{2\left(\tan^2(\theta)+2\right)}{3r\sin(\theta)}\mathrm{d}\theta^{-1} \tag{66}$$

$$\mathrm{d}y^{-1} = \frac{2\cot(\theta)-1}{3}\mathrm{d}r^{-1} + \frac{2\left(\cot^2(\theta)+2\right)}{3r\cos(\theta)}\mathrm{d}\theta^{-1} \tag{67}$$

To construct a covariant fractional derivative the restriction $\mathrm{Re}(\nu) \geq 0$ is made. A covariant fractional derivative must have the following properties.

$$\lim_{\nu\to 1} {}^{\nu}_{\tilde{a}_i}\nabla_{y_i} = \nabla_{y_i} \tag{68}$$

$$\lim_{\{y\}\to\{x\}} {}^{\nu}_{\tilde{a}_i}\nabla_{y_i} = {}_{a_i}D^{\nu}_{x_i} \tag{69}$$

$$\sum_{i=1}^{n} J^i_j(y,z,\nu)\, {}^{\nu}_{\tilde{a}_i}\nabla_{y_i} = {}^{\nu}_{\tilde{\tilde{a}}_j}\nabla_{z_j} \tag{70}$$

Equation (68) is merely a statement that we get the covariant derivative that we are familiar with from differential geometry; i.e. covariantly constant metric and parallel transport. Please note that the metric need only be covariantly constant for the case of derivatives of order 1 and for a background space of order 1. This is because parallel transport is tied to the notion of the tangent vector. Equation (69) says that when the curvilinear coordinates become Cartesian the covariant fractional derivative becomes an ordinary fractional derivative. Just as in differential geometry the covariant derivative becomes a partial derivative when space becomes flat and the coordinates Cartesian. Equation (70) is just the statement that the fractional covariant derivative transform as a vector.

From FDF I the following transformation rules are available,

$$ {}_{a_j}\mathrm{D}^{\nu}_{x_j} = \sum_{l=1}^{n} \mathrm{J}^{l}{}_{j}(y,x,\nu)\, {}_{\tilde{a}_l}\mathrm{D}^{\nu}_{y_l}\,, \tag{71} $$

$$ \mathrm{V}_k(x) = \sum_{a=1}^{n} \mathrm{J}^{a}{}_{k}(y,x,\nu)\mathrm{V}_a(y). \tag{72} $$

The covariant derivative of a vector must transform as,

$$ \sum_{i,m=1}^{n} \mathrm{J}^{i}{}_{j}(y,x,\nu)\mathrm{J}^{m}{}_{k}(y,x,\nu)\left({}^{\nu}_{\tilde{a}_i}\nabla_{y_i}\mathrm{V}_m(y)\right) = {}_{a_i}\mathrm{D}^{\nu}_{x_j}\mathrm{V}_k(x). \tag{73} $$

Now substitute the known transformation rules into equation (73).

$$ \sum_{i,m=1}^{n} \mathrm{J}^{i}{}_{j}(y,x,\nu)\mathrm{J}^{m}{}_{k}(y,x,\nu)\left({}^{\nu}_{\tilde{a}_i}\nabla_{y_i}\mathrm{V}_m(y)\right) = \sum_{a,l=1}^{n} \mathrm{J}^{l}{}_{j}(y,x,\nu)\, {}_{\tilde{a}_l}\mathrm{D}^{\nu}_{y_l}\left(\mathrm{J}^{a}{}_{k}(y,x,\nu)\mathrm{V}_a(y)\right) \tag{74} $$

Multiply equation (74) by $\sum_{j=1}^{n} \mathrm{J}_b^{\ j}(x,y,\nu)$ and recall that, $\sum_{j=1}^{n} \mathrm{J}^i_{\ j}(y,x,\nu)\mathrm{J}_b^{\ j}(x,y,\nu) = \delta_b^i$ gives,

$$\sum_{m=1}^{n} \mathrm{J}^m_{\ k}(y,x,\nu)\left({}_{\tilde{a}_b}^{\ \nu}\nabla_{y_b} \mathrm{V}_m(y)\right) = \sum_{a=1}^{n} {}_{\tilde{a}_b}\mathrm{D}_{y_b}^{\nu}\left(\mathrm{J}^a_{\ k}(y,x,\nu)\mathrm{V}_a(y)\right). \tag{75}$$

Multiply equation (75) by $\sum_{k=1}^{n} \mathrm{J}_l^{\ k}(x,y,\nu)$, simplify again, and the fractional covariant derivative is seen to be

$$ {}_{\tilde{a}_b}^{\ \nu}\nabla_{y_b} \mathrm{V}_l(y) = \sum_{k,a=1}^{n} \mathrm{J}_l^{\ k}(x,y,\nu)\, {}_{\tilde{a}_b}\mathrm{D}_{y_b}^{\nu}\left(\mathrm{J}^a_{\ k}(y,x,\nu)\mathrm{V}_a(y)\right). \tag{76}$$

If $\nu \to 1$ then the expression reduces to,

$$\nabla_b V_l = \sum_{k,a=1}^{n} J_l^k\left(\partial_{y_b} V_a J_k^a + V_a \partial_{y_b} J_k^a\right) = \partial_{y_b} V_l + \sum_{k,a=1}^{n} V_a J_l^k \partial_{y_b} J_k^a . \tag{77}$$

Equation (77) is the usual expression for a covariant derivative, written in terms of the coordinate transformation matrix.

Now expand equation (76) using the product rule for fractional derivatives

$$ {}_{\tilde{a}_b}^{\ \nu}\nabla_{y_b} \mathrm{V}_l(y) = \sum_{k,a=1}^{n} \mathrm{J}_l^{\ k}(x,y,\nu) \sum_{s=0}^{\infty} \binom{\nu}{s} \left({}_{\tilde{a}_b}\mathrm{D}_{y_b}^{\nu-s} \mathrm{V}_a(y)\right)\left(\partial_{y_b}^{s} \mathrm{J}^a_{\ k}(y,x,\nu)\right). \tag{78}$$

If the first term of the infinite series is separated out the connection can be isolated.

$$ {}_{\tilde{a}_b}^{\ \nu}\nabla_{y_b} \mathrm{V}_l(y) = \sum_{k,a=1}^{n} \mathrm{J}_l^{\ k}(x,y,\nu)\left\{\left({}_{\tilde{a}_b}\mathrm{D}_{y_b}^{\nu}\mathrm{V}_a(y)\right)\left(\mathrm{J}^a{}_k(y,x,\nu)\right) + \sum_{s=1}^{\infty}\binom{\nu}{s}\left({}_{\tilde{a}_b}\mathrm{D}_{y_b}^{\nu-s}\mathrm{V}_a(y)\right)\left(\partial_{y_b}^{s}\mathrm{J}^a{}_k(y,x,\nu)\right)\right\} \tag{79} $$

$$ {}_{\tilde{a}_b}^{\ \nu}\nabla_{y_b} \mathrm{V}_l(y) = {}_{\tilde{a}_b}\mathrm{D}_{y_b}^{\nu}\mathrm{V}_l(y) + \sum_{k,a=1}^{n} \mathrm{J}_l^{\ k}(x,y,\nu)\sum_{s=1}^{\infty}\binom{\nu}{s}\left({}_{\tilde{a}_b}\mathrm{D}_{y_b}^{\nu-s}\mathrm{V}_a(y)\right)\left(\partial_{y_b}^{s}\mathrm{J}^a{}_k(y,x,\nu)\right) \tag{80} $$

This is somewhat more complicated than the usual connection coefficients that arise when constructing a covariant derivative. So, adopt the following notation and refer to the connection as a connection functional (due to the integrals involved in the sum).

$$ {}_{\tilde{a}_b}^{\ \nu}\nabla_{y_b} \mathrm{V}_l(y) = {}_{\tilde{a}_b}\mathrm{D}_{y_b}^{\nu}\mathrm{V}_l(y) + {}_{\tilde{a}_b}^{\ \nu}\gamma_{lb}(\mathrm{V}) \tag{81} $$

$$ {}_{\tilde{a}_b}^{\ \nu}\gamma_{lb}(\mathrm{V}) = \sum_{k,a=1}^{n} \mathrm{J}_l^{\ k}(x,y,\nu)\sum_{s=1}^{\infty}\binom{\nu}{s}\left(\partial_{y_b}^{s}\mathrm{J}^a{}_k(y,x,\nu)\left({}_{\tilde{a}_b}\mathrm{D}_{y_b}^{\nu-s}\mathrm{V}_a(y)\right)\right) \tag{82} $$

As a cautionary note, recall that these objects are defined on a Euclidean space, i.e. there is no curvature. For a curved manifold there would be difficult issues concerning the non-local nature of fractional derivatives. Another difficult problem is to find an expression for a covariant derivative that is a different order than the background space. How does one construct ${}_{P}^{\mu}\nabla_{y_a}$ on a space where the vectors transform under $J_b^a(y,z,\nu)$? The obvious path to approach this problem would be to use the chain rule. However, the chain rule in fractional calculus is computationally difficult to use for an arbitrary order (see pages 97 and 98 of [21]). For an alternate discussion of the fractional covariant derivative see Kobelev [11].

## IV. Matrix Order Forms.

In the appendix fractional derivatives are extended to matrix order derivatives. Matrix order derivatives are shown to be well defined for all square matrices over the complex numbers ($C^{m\times m}$). Consider the definition of a fractional exterior differintegral.

$$\mathrm{d}^{\nu} = \sum_{i=1}^{n} \mathrm{d}x_i^{\nu}\ {}_{a_i}\mathrm{D}_{x_i}^{\nu} \qquad (83)$$

To construct a matrix order form, replace the parameter $\nu \in C$ by a matrix $\mathrm{A} \in C^{m\times m}$. This can be expressed directly or, using the spectral theorem (see page 517 of [15]).

$$\mathrm{d}^{\mathrm{A}} = \sum_{i=1}^{n} \mathrm{d}x_i^{\mathrm{A}}\ {}_{a_i}\mathrm{D}_{x_i}^{\mathrm{A}} = \sum_{i=1}^{n} \mathrm{d}x_i^{\mathrm{A}} \mathrm{P} \begin{pmatrix} {}_{a_i}\mathrm{D}_{x_i}^{\lambda_1} & & \\ & \mathrm{O} & \\ & & {}_{a_i}\mathrm{D}_{x_i}^{\lambda_m} \end{pmatrix} \mathrm{P}^{-1} \qquad (84)$$

$$\mathrm{d}^{\mathrm{A}} = \sum_{i=1}^{n} \mathrm{d}x_i^{\mathrm{A}} \sum_{j=1}^{k} G_j\ {}_{a_i}\mathrm{D}_{x_i}^{\lambda_j} \qquad (85)$$

Where $\mathrm{A} = \mathrm{PDP}^{-1}$, D is a diagonal matrix whose entries are the eigenvalues of A, $\mathrm{A} = \sum_{i=1}^{k} \mathrm{G}_i \lambda_i$, k is the number of unrepeated eigenvalues, $\lambda_i$ are the eigenvalues, and the $\mathrm{G}_i$ are the spectral projectors.

To develop an understanding of these new objects, consider the coordinate transformation rules for matrix order forms.

$$\mathrm{d}x^{\mathrm{A}}_{x_i} = \sum_{j=1}^{n} \mathrm{d}y^{\mathrm{A}}_{j} \mathrm{J}^{i}_{j}(\mathrm{A},x,y) \tag{86}$$

For now assume that A is diagonalizable, the Jordan case will be dealt with in a later paper. Express the Cartesian coordinates in terms of the curvilinear coordinates, $x_i = x_i(y)$, and compute the exterior fractional differintegral in the two coordinate systems.

$$\mathrm{d}^{\mathrm{A}}(x_k) = \mathrm{d}^{\mathrm{A}}(x_k(y)) \tag{87}$$

$$\sum_{i=1}^{n} \mathrm{d}x_i^{\mathrm{A}} \, {}_{a_i}\mathrm{D}^{\mathrm{A}}_{x_i}(x_k) = \sum_{j=1}^{n} \mathrm{d}y_j^{\mathrm{A}} \, {}_{\tilde{a}_j}\mathrm{D}^{\mathrm{A}}_{y_j}(x_k(y)) \tag{88}$$

To find the coordinate transformation matrix equation (88) needs to be solved for the $\mathrm{d}x_i^{\mathrm{A}}$. ${}_{a_i}\mathrm{D}^{\mathrm{A}}_{x_i}(x_k)$ and ${}_{\tilde{a}_j}\mathrm{D}^{\mathrm{A}}_{y_j}(x_k(y))$ form $n \times n$ matrices as $i,j,k \in \{1, \ldots, n\}$. Hence equation (88) represents $n^2$ equations for the n $\mathrm{d}x_i^{\mathrm{A}}$. This is an over determined system, hence, $\mathrm{d}x_i^{\mathrm{A}}$ cannot be viewed as a single object. The simplest thing to do is to view $\mathrm{d}x_i^{\nu}$ as an object continually depending on the parameter $\nu$, and then write (see page 526 of [15]),

$$\mathrm{d}x_i^{\mathrm{A}} = \mathrm{P} \begin{pmatrix} \mathrm{d}x_i^{\lambda_1} & & \\ & \ddots & \\ & & \mathrm{d}x_i^{\lambda_m} \end{pmatrix} \mathrm{P}^{-1}. \tag{89}$$

With this assumption consider equation (88)

$$\sum_{i=1}^{n} \mathrm{P} \begin{pmatrix} \mathrm{d}x_i^{\lambda_1} & & \\ & \ddots & \\ & & \mathrm{d}x_i^{\lambda_m} \end{pmatrix} \mathrm{P}^{-1} {}_{a_i}\mathrm{D}_{x_i}^{\mathrm{A}}(x_k) = \sum_{j=1}^{n} \mathrm{P} \begin{pmatrix} \mathrm{d}y_j^{\lambda_1} & & \\ & \ddots & \\ & & \mathrm{d}y_j^{\lambda_m} \end{pmatrix} \mathrm{P}^{-1} {}_{\tilde{a}_j}\mathrm{D}_{y_j}^{\mathrm{A}}(x_k(y)). \tag{90}$$

Equation (90) can be simplified if ${}_{a_i}\mathrm{D}_{x_i}^{\mathrm{A}}$ and ${}_{\tilde{a}_j}\mathrm{D}_{y_j}^{\mathrm{A}}$ are expressed in terms of their diagonal matrices and all factors of $\mathrm{P}^{-1}$ and P are either multiplied together or factored out.

$$\sum_{i=1}^{n} \begin{pmatrix} \mathrm{d}x_i^{\lambda_1}\, {}_{a_i}\mathrm{D}_{x_i}^{\lambda_1}(x_k) & & \\ & \ddots & \\ & & \mathrm{d}x_i^{\lambda_m}\, {}_{a_i}\mathrm{D}_{x_i}^{\lambda_m}(x_k) \end{pmatrix} = \sum_{j=1}^{n} \begin{pmatrix} \mathrm{d}y_j^{\lambda_1}\, {}_{\tilde{a}_j}\mathrm{D}_{y_j}^{\lambda_1}(x_k(y)) & & \\ & \ddots & \\ & & \mathrm{d}y_j^{\lambda_m}\, {}_{\tilde{a}_j}\mathrm{D}_{y_j}^{\lambda_m}(x_k(y)) \end{pmatrix} \tag{91}$$

This can be expressed as a single equation,

$$\sum_{i=1}^{n} \mathrm{d}x_i^{\lambda_l}\, {}_{a_i}\mathrm{D}_{x_i}^{\lambda_l}(x_k) = \sum_{j=1}^{n} \mathrm{d}y_j^{\lambda_l}\, {}_{\tilde{a}_j}\mathrm{D}_{y_j}^{\lambda_l}(x_k(y)), \tag{92}$$

where $l \in \{1, \ldots, m\}$. A coordinate transformation matrix can be constructed using Cramer's rule as was done in section 4. If all the eigenvalues are such that $\mathrm{Re}(\lambda_l) \geq 0$ then the construction from FDF I can be used. In either case a prescription is available for the construction of the coordinate transformation matrix for matrix order forms.

$$\begin{pmatrix} \mathrm{d}x_i^{\lambda_1} & & \\ & \ddots & \\ & & \mathrm{d}x_i^{\lambda_m} \end{pmatrix} = \sum_{j=1}^{n} \begin{pmatrix} \mathrm{d}y_j^{\lambda_1}\, {}_{\tilde{a}_j}\mathrm{J}_j^i(\lambda_1, x, y) & & \\ & \ddots & \\ & & \mathrm{d}y_j^{\lambda_m}\, \mathrm{J}_j^i(\lambda_m, x, y) \end{pmatrix}. \tag{93}$$

If P and $\mathrm{P}^{-1}$ are inserted in the appropriate places equation (93) becomes,

$$dx^{A}_{x_i} = \sum_{j=1}^{n} dy^{A}_{j} J^{i}_{j}(A,x,y) \qquad (94)$$

where,

$$J^{i}_{j}(A,x,y) = P \begin{pmatrix} J^{i}_{j}(\lambda_1,x,y) & & \\ & \circ & \\ & & J^{i}_{j}(\lambda_m,x,y) \end{pmatrix} P^{-1}. \qquad (95)$$

With the coordinate transformation matrix found a metric and line element can be constructed in the same manor as in FDF I. A matrix order covariant derivative can also be constructed provided that attention is restricted to matrix orders that are positive definite.

$${}^{A}_{\tilde{a}_i}\nabla_{y_b} V_l(y) = {}_{\tilde{a}_i} D^{A}_{y_b} V_l(y) + {}^{A}_{\tilde{a}_i}\gamma_{lb}(V) \qquad (96)$$

$${}^{A}_{P}\gamma_{lb}(V) = \sum_{k,a=1}^{n} J_l{}^{k}(x,y,A) \sum_{s=1}^{\infty} \begin{pmatrix} A \\ sI \end{pmatrix} \left({}_{P}D^{A-sI}_{y_b} V_a(y)\right)\left(\partial^{sI}_{y_b} J^{a}{}_{k}(y,x,A)\right) \qquad (97)$$

$J_l{}^{k}(A,x,y)$ is defined in equation (95), the other components of equation (97) are defined in equations (98), (99), and (100).

$$\begin{pmatrix} A \\ sI \end{pmatrix} = P \begin{pmatrix} \begin{pmatrix} \lambda_1 \\ s \end{pmatrix} & & \\ & \circ & \\ & & \begin{pmatrix} \lambda_m \\ s \end{pmatrix} \end{pmatrix} P^{-1} \qquad (98)$$

$$
{}_{\tilde{a}_i}\mathrm{D}_{y_b}^{\mathrm{A}-s\mathrm{I}} = \mathrm{P}\begin{pmatrix} {}_{\tilde{a}_i}\mathrm{D}_{y_b}^{\lambda_1 - s} & & \\ & \ddots & \\ & & {}_{\tilde{a}_i}\mathrm{D}_{y_b}^{\lambda_\mathrm{m} - s} \end{pmatrix}\mathrm{P}^{-1} \tag{99}
$$

$$
\partial_{y_b}^{\mathrm{sI}} = \begin{pmatrix} \partial_{y_b}^{\mathrm{s}} & & \\ & \ddots & \\ & & \partial_{y_b}^{\mathrm{s}} \end{pmatrix} \tag{100}
$$

I is the $m \times m$ identity matrix. Equations (95)-(99) can also be expressed using the spectral theorem.

The Poincaré lemma also holds true for matrix order exterior differintegrals. Let $\alpha \in \mathrm{G}(\nu_1,\ldots\ \nu_1,\nu_2,\ldots\ ,\nu_2,\ldots\ ,\nu_k,\ldots\ \nu_k,\mathrm{n})$ and allow the coefficients of $\alpha$ to possibly be matrix valued, then $\mathrm{d}^{\mathrm{A}}\mathrm{d}^{\mathrm{A}}\alpha$ can be written as,

$$
\mathrm{d}^{\mathrm{A}}\mathrm{d}^{\mathrm{A}}\alpha = \sum_{i=1}^{n}\sum_{j=1}^{n}\mathrm{P}\begin{pmatrix} \mathrm{d}x_i^{\lambda_1} \wedge \mathrm{d}x_j^{\lambda_1}\ {}_{a_i}\mathrm{D}_{x_i}^{\lambda_1}\ {}_{a_j}\mathrm{D}_{x_j}^{\lambda_1} & & \\ & \ddots & \\ & & \mathrm{d}x_i^{\lambda_m} \wedge \mathrm{d}x_j^{\lambda_m}\ {}_{a_i}\mathrm{D}_{x_i}^{\lambda_\mathrm{m}}\ {}_{a_j}\mathrm{D}_{x_j}^{\lambda_\mathrm{m}} \end{pmatrix}\mathrm{P}^{-1}\alpha. \tag{101}
$$

Each component of the right-hand side of equation (101) is now equivalent to the problem considered in section 3. Hence, $\mathrm{d}^{\mathrm{A}}\mathrm{d}^{\mathrm{A}}\ \alpha = 0$ for any $\mathrm{A} \in C^{m\times m}$ that is diagonalizable and for any $\alpha \in \mathrm{F}(\nu,\mathrm{m},\mathrm{n})$ (assuming that the components of $\alpha$ are differintegrable to the appropriate orders). For matrices that are only Jordan diagonalizable the result does not hold. This is because the upper triangular Jordan blocks do not commute.

## V. Conclusion.

In this paper the results of FDF I were extended to include an inner product, Hodge dual, and covariant derivative for fractional differential forms. The connection for covariant fractional order derivatives was found to be somewhat more complex than for covariant derivatives of order one. The resulting formula does reduce to the usual formula from differential geometry. The notion of a fractional differential form was also extended to fractional integral and matrix order forms. Coordinate transformation rules were worked out for both of these objects with a specific example presented for integral order forms. Matrix order forms were constructed by combining matrix order fractional calculus (see the appendix) with exterior derivatives. This was done in the same way that fractional order derivatives were combined with exterior derivatives in FDF I. The Poincaré lemma was found to be true for all orders of exterior differintegrals provided that the order of the differintegral was a diagonalizable matrix or any complex number.

One final note, the Riemann-Liouville fractional differintegral was used for the construction of fractional forms. There is no reason that the Grünwald-Letnikov, Weyl, or the Caputo definition of fractional differintegrals could not be used to define fractional forms. See references [17] and [21] for the definitions of these other fractional differintegrals.

**Acknowledgment**: The authors would like to thank P. Dorcey for helpful comments and a critical reading of the paper and Jean Krisch for asking the right questions to get all of this started.

**Appendix**: **Matrix Order Differintegration**

In this appendix the formulae from the Riemann-Liouville formulation of fractional calculus are adapted to matrix order for any matrix $\mathrm{A} \in C^{n\times n}$ ($n \times n$ square matrices over the complex numbers). A brief review of matrix properties is provided to set notation and to provide a convenient reference for the reader.

The concepts of integration and differentiation have been expanded many times throughout the development of calculus. Almost immediately after the formulation of classical calculus by Leibnitz and Newton the question of half order derivatives arose (see [16] and [17] for historical reviews). Derivatives of complex and purely imaginary order were later developed (see e.g. [13] and references there in). Phillips considered fractional order derivatives of matrix-valued functions with respect to their matrix arguments, fractional matrix calculus. These operations were found to be well defined and subsequently applied to econometric distribution theory (see [18]-[20]).

To develop the notion of matrix order derivatives and integrals (differintegrals) it is useful to review the properties of functions with matrix order arguments. Three different means of computing functions with matrix order arguments will be used in this paper. The method chosen will be determined by the properties of the matrix being used and the specific application being considered.

A given non-zero matrix $\mathrm{A} \in C^{m\times m}$ can be placed into one or more of the following sets; Jordan block diagonalizable, diagonalizable, and/or normal. These sets will be denoted by $\mathrm{B}_m$, $\mathrm{D}_m$, and $\mathrm{N}_m$, respectively. Note that $\mathrm{N}_m \subset \mathrm{D}_m \subset \mathrm{B}_m$ and every non-zero matrix $\mathrm{A} \in C^{m\times m}$ is at least Jordan block diagonalizable. Recall that the set of normal matrices contains the following subsets: real symmetric, real skew-symmetric, positive and negative definite, Hermitian, skew-Hermitian, orthogonal, and unitary (see page 548 of [15] for further details). If a matrix A is normal then there exists a unitary matrix U such that

$$\mathrm{A} = \mathrm{UDU}^{*}. \tag{102}$$

Where D is a diagonal matrix with the eigenvalues of A as the entries and $*$ denotes conjugate transpose. If A is real and normal then equation (102) reduces to

$$\mathrm{A} = \mathrm{ODO}^{\mathrm{T}} \tag{103}$$

and O is an orthogonal matrix. If a matrix A is diagonalizable then there exists an invertible matrix P such that

$$\mathrm{A} = \mathrm{PDP}^{-1}. \tag{104}$$

Alternatively, if A is diagonalizable the spectral theorem (see page 517 of [15]) can be used to find another representation for A. If $\mathrm{A} \in \mathrm{D}_m$ then there exists a set of matrices $\{\mathrm{G}_1, \ldots, \mathrm{G}_k\}$ such that,

$$A = \sum_{i=1}^{k} G_i \, \lambda_i \,, \tag{105}$$

where $\{\lambda_1, \ldots, \lambda_k\}$ are the unrepeated eigenvalues of A. The matrices $\{\mathrm{G}_1, \ldots, \mathrm{G}_k\}$ have the following properties;

$$\mathrm{G}_i \mathrm{G}_j = 0 \quad \text{for} \quad i \neq j \,, \tag{106}$$

$$G_i G_i = G_i \quad , \tag{107}$$

$$I = \sum_{i=1}^{k} G_i \quad . \tag{108}$$

If a matrix is Jordan block diagonalizable, then there is an invertible matrix P such that

$$A \;=\; PJP^{-1} \;=\; P \begin{pmatrix} J(\lambda_1) & & 0 \\ & \ddots & \\ 0 & & J(\lambda_k) \end{pmatrix} P^{-1} \tag{109}$$

Where $\lambda_1, \ldots, \lambda_k$ are the distinct eigenvalues of A ($k < m$), and $J(\lambda_j)$ is the Jordan segment for the eigenvalue $\lambda_j$.

$$J(\lambda_j) = \begin{pmatrix} \lambda_j & 1 & & \\ & \ddots & \ddots & \\ & & \ddots & 1 \\ & & & \lambda_j \end{pmatrix} \tag{110}$$

There are several ways to define a matrix function. Let $A \in N_m$ with eigenvalues $\lambda_1, \ldots, \lambda_m$ (they need not be distinct) and let g(z) be a function that is defined for all the eigenvalues of A. Then g(A) can be expressed as

$$g(A) = U \begin{pmatrix} g(\lambda_1) & & \\ & \ddots & \\ & & g(\lambda_m) \end{pmatrix} U^* . \tag{111}$$

If A is real replace U by O and $U^*$ by $O^T$. If A is diagonalizable then equation (111) would be written as,

$$g(A) = P\begin{pmatrix} g(\lambda_1) & & \\ & \ddots & \\ & & g(\lambda_m) \end{pmatrix}P^{-1}. \tag{112}$$

The spectral representation can also be used (in either case).

$$g(A) = \sum_{i=1}^{k} G_i \, g(\lambda_i) \tag{113}$$

Where $\lambda_1, \ldots, \lambda_k$ are the distinct eigenvalues. For matrices that are Jordan block diagonalizable but not in $D_m$ the situation is somewhat more complicated.

$$g(A) = P\begin{pmatrix} g(J(\lambda_1)) & & \\ & \ddots & \\ & & g(J(\lambda_k)) \end{pmatrix}P^{-1} \tag{114}$$

Where,

$$g(J(\lambda_i)) = \begin{pmatrix} g(\lambda_i) & g'(\lambda_i) & g''(\lambda_i)/2! & \cdots & g^{(l-1)}(\lambda_i)/(l-1)! \\ & g(\lambda_i) & g'(\lambda_i) & \cdots & g^{(l-2)}(\lambda_i)/(l-2)! \\ & & \ddots & \ddots & \vdots \\ & & & g(\lambda_i) & g'(\lambda_i) \\ & & & & g(\lambda_i) \end{pmatrix}. \tag{115}$$

$g(J(\lambda_i))$ is an $l \times l$ upper triangular matrix. Note that for this case the function need be at least differentiable of order $l-1$ at $\lambda_i$. For further details see [15].

Using equations (4) and (5) of section 2 an analytic function can be defined (see page 49 of [17]) for a given f(x) that is suitably differentiable.

$$g(\lambda) = {}_a\mathrm{D}_x^{\lambda}\mathrm{f}(x) = \begin{cases} \dfrac{1}{\Gamma(-\lambda)}\displaystyle\int_a^x \frac{\mathrm{f}(\xi)\mathrm{d}\xi}{(x-\xi)^{1+\lambda}}\ , & \mathrm{Re}(\lambda) < 0, \\ \dfrac{\partial^n}{\partial x^n}\left[\dfrac{1}{\Gamma(n-\lambda)}\displaystyle\int_a^x \frac{\mathrm{f}(\xi)\mathrm{d}\xi}{(x-\xi)^{\lambda-n+1}}\right], & \dfrac{\mathrm{Re}(\lambda) \geq 0}{n > \mathrm{Re}(\lambda)\ (\text{n is whole})} \end{cases} \tag{116}$$

Let $\mathrm{A} \in \mathrm{N}_m$ then the differintegral of order A is given by,

$${}_a\mathrm{D}_x^A = \mathrm{U}\begin{pmatrix} {}_a\mathrm{D}_x^{\lambda_1} & & \\ & \circ & \\ & & {}_a\mathrm{D}_x^{\lambda_m} \end{pmatrix}\mathrm{U}^*. \tag{117}$$

If A is real and normal then,

$${}_a\mathrm{D}_x^A = \mathrm{O}\begin{pmatrix} {}_a\mathrm{D}_x^{\lambda_1} & & \\ & \circ & \\ & & {}_a\mathrm{D}_x^{\lambda_m} \end{pmatrix}\mathrm{O}^{\mathrm{T}} \tag{118}$$

If $\mathrm{A} \in \mathrm{D}_m$ then

$${}_a\mathrm{D}_x^A = \mathrm{P}\begin{pmatrix} {}_a\mathrm{D}_x^{\lambda_1} & & \\ & \circ & \\ & & {}_a\mathrm{D}_x^{\lambda_m} \end{pmatrix}\mathrm{P}^{-1} \tag{119}$$

Any of the above three cases, equations (117), (118), or (119), can be expressed using the spectral theorem.

$$ {}_a\mathrm{D}_x^{A} = \sum_{i=1}^{k} G_i \; {}_a\mathrm{D}_x^{\lambda_i} \qquad (120) $$

Where the $G_i$ are the matrices as given by the spectral theorem and the sum is over the unrepeated eigenvalues.

For matrices that are Jordan block diagonalizable but not diagonalizable some notation must be introduced. Recall equation (116).

$$ g(\lambda) = {}_a\mathrm{D}_x^{\lambda}\mathrm{f}(x) = \begin{cases} \dfrac{1}{\Gamma(-\lambda)} \displaystyle\int_a^x \frac{\mathrm{f}(\xi)\mathrm{d}\xi}{(x-\xi)^{1+\lambda}} \;, & \mathrm{Re}(\lambda) < 0, \\ \dfrac{\partial^n}{\partial x^n} \left[ \dfrac{1}{\Gamma(n-\lambda)} \displaystyle\int_a^x \frac{\mathrm{f}(\xi)\mathrm{d}\xi}{(x-\xi)^{\lambda-n+1}} \right], & \dfrac{\mathrm{Re}(\lambda) \ge 0}{\mathrm{n} > \mathrm{Re}(\lambda) \text{ (n is whole)}} \end{cases} \qquad (121) $$

Derivatives of $g(\lambda)$ with respect to $\lambda$ will appear in the expression for the Jordan matrices.

$$ \frac{\mathrm{d}^k}{\mathrm{d}\lambda^k} g(\lambda) = {}_a^k\mathrm{D}_x^{\lambda}\mathrm{f}(x) = \begin{cases} \dfrac{\mathrm{d}^k}{\mathrm{d}\lambda^k} \dfrac{1}{\Gamma(\lambda)} \displaystyle\int_a^x \frac{\mathrm{f}(\xi)\mathrm{d}\xi}{(x-\xi)^{1-\lambda}} \;, & \mathrm{Re}(\lambda) < 0, \\ \dfrac{\mathrm{d}^k}{\mathrm{d}\lambda^k} \dfrac{\partial^n}{\partial x^n} \left[ \dfrac{1}{\Gamma(n-\lambda)} \displaystyle\int_a^x \frac{\mathrm{f}(\xi)\mathrm{d}\xi}{(x-\xi)^{\lambda-n+1}} \right] & \dfrac{\mathrm{Re}(\lambda) \ge 0}{\mathrm{n} > \mathrm{Re}(\lambda) \text{ (n is whole)}} \end{cases} \qquad (122) $$

In equation (122) k is restricted to be a whole number. The ${}_a^k\mathrm{D}_x^{\lambda}$ can be expressed in terms of ${}_a\mathrm{D}_x^{\lambda}$. For example, if k = 1 the following is obtained.

$$ {}_{a}^{1}\mathrm{D}_x^{\lambda} f(x) = -\left(\frac{d}{d\lambda}\ln(-\lambda) + \ln(x)\right){}_{a}\mathrm{D}_x^{\lambda} f(x) - \sum_{j=1}^{\infty} \frac{{}_{a}\mathrm{D}_x^{\lambda}(x^j f(x))}{j\, x^j} \qquad \mathrm{Re}(\lambda) < 0 \qquad (123) $$

$$ {}_{a}^{1}\mathrm{D}_x^{\lambda} f(x) = -\frac{\partial^n}{\partial x^n}\left(\left(\frac{d}{d\lambda}\ln(-\lambda) + \ln(x)\right){}_{a}\mathrm{D}_x^{\lambda-n} f(x) + \sum_{j=1}^{\infty} \frac{{}_{a}\mathrm{D}_x^{\lambda-j}(x^j f(x))}{j\, x^j}\right) \frac{\mathrm{Re}(\lambda)\ \ge\ 0}{\mathrm{n} > \mathrm{Re}(\lambda)\ (\text{n is whole})} \quad (124) $$

Similar expressions are available for the higher order derivatives of ${}_{a}\mathrm{D}_x^{\lambda}$. With this notation matrix order differintegrals for $\mathrm{A} \in \mathrm{B}_m$ are denoted by

$$ {}_{a}\mathrm{D}_x^{A} = \mathrm{P}\begin{pmatrix} {}_{a}\mathrm{D}_x^{J(\lambda_1)} & & \\ & \ddots & \\ & & {}_{a}\mathrm{D}_x^{J(\lambda_k)} \end{pmatrix}\mathrm{P}^{-1} \qquad (125) $$

where,

$$ {}_{a}\mathrm{D}_x^{J(\lambda_i)} = \begin{pmatrix} {}_{a}^{0}\mathrm{D}_x^{\lambda_i} & {}_{a}^{1}\mathrm{D}_x^{\lambda_i}/1! & \cdots & {}_{a}^{l\text{-}1}\mathrm{D}_x^{\lambda_i}/(l-1)! \\ & {}_{a}^{0}\mathrm{D}_x^{\lambda_i} & \cdots & {}_{a}^{l\text{-}2}\mathrm{D}_x^{\lambda_i}/(l-2)! \\ & & \ddots & \vdots \\ & & & {}_{a}^{0}\mathrm{D}_x^{\lambda_i} \end{pmatrix} \qquad (126) $$

To determine some of the properties of matrix order differintegrals consider the composition of two matrix order differintegrals. Let A and B $\in \mathrm{D}_m$ such that $\mathrm{A} = \mathrm{PDP}^{-1}$ and $\mathrm{B} = \mathrm{QEQ}^{-1}$ where D and E are diagonal matrices with the eigenvalues of A and B as entries. Denote the eigenvalues of A as $\lambda_i$ and the eigenvalues of B as $\rho_i$. Then,

$$ {}_aD_x^A\,{}_aD_x^B = P\begin{pmatrix} {}_aD_x^{\lambda_1} & & \\ & \circ & \\ & & {}_aD_x^{\lambda_m} \end{pmatrix}P^{-1}Q\begin{pmatrix} {}_aD_x^{\rho_1} & & \\ & \circ & \\ & & {}_aD_x^{\rho_m} \end{pmatrix}Q^{-1}. \qquad (127) $$

To simplify this denote by $R = P^{-1}Q$ and the components of R as $R_{ij}$. Equation (127) now has the form,

$$ {}_aD_x^A\,{}_aD_x^B = P\left[R_{ij}\,{}_aD_x^{\lambda_i}\,{}_aD_x^{\rho_j}\right]Q^{-1}. \qquad (128) $$

There is no sum over the repeated indices, they merely denote the components of the matrix between P and $Q^{-1}$. Alternatively the spectral theorem can be used to obtain another representation for the right hand side of equation (127).

$$ {}_aD_x^A\,{}_aD_x^B = \sum_{i=1}^{k} G_i\,{}_aD_x^{\lambda_i}\sum_{j=1}^{l} H_j\,{}_aD_x^{\rho_j} = \sum_{i=1}^{k}\sum_{j=1}^{l} G_i H_j\,{}_aD_x^{\lambda_i}\,{}_aD_x^{\rho_j} \qquad (129) $$

Where $A = \sum_{i=1}^{k} G_i\,\lambda_i$ and $B = \sum_{j=1}^{l} H_j\,\rho_j$.

Equation (4) carries over to the matrix order case.

$$ \frac{\partial^m}{\partial x^m} D_x^A f(x) = \frac{\partial^m}{\partial x^m}\sum_{i=1}^{k} G_i\,{}_aD_x^{\lambda_i} f(x) \qquad (130) $$

$$ = \sum_{i=1}^{k} G_i\,\frac{\partial^m}{\partial x^m}\,{}_aD_x^{\lambda_i} f(x) \qquad (131) $$

$$= \sum_{i=1}^{k} G_i \, {}_a\mathrm{D}_x^{\lambda_i + m} \mathrm{f}(x) \tag{132}$$

$$= {}_a\mathrm{D}_x^{A+m\mathrm{I}} \mathrm{f}(x) \tag{133}$$

I is the identity matrix, m is a whole number, and f(x) may be a scalar, vector, or matrix order function.

Equation (7) carries over in the following sense, Let A be a matrix such that $\mathrm{Re}(\lambda_i) \geq 0$. Now consider equation (127) with B replaced by –A.

$${}_a\mathrm{D}_x^{A} \, {}_a\mathrm{D}_x^{-A} = \mathrm{P} \begin{pmatrix} {}_a\mathrm{D}_x^{\lambda_1} & & \\ & \mathrm{O} & \\ & & {}_a\mathrm{D}_x^{\lambda_m} \end{pmatrix} \begin{pmatrix} {}_a\mathrm{D}_x^{-\lambda_1} & & \\ & \mathrm{O} & \\ & & {}_a\mathrm{D}_x^{-\lambda_m} \end{pmatrix} \mathrm{P}^{-1} \tag{134}$$

$${}_a\mathrm{D}_x^{A} \, {}_a\mathrm{D}_x^{-A} = \mathrm{P} \begin{pmatrix} {}_a\mathrm{D}_x^{\lambda_1} \, {}_a\mathrm{D}_x^{-\lambda_1} & & \\ & \mathrm{O} & \\ & & {}_a\mathrm{D}_x^{\lambda_m} \, {}_a\mathrm{D}_x^{-\lambda_m} \end{pmatrix} \mathrm{P}^{-1} \tag{135}$$

Which, by equation (7), reduces to the identity operator.

Suppose that A and B commute (and are diagonalizable), then A and B can be diagonalized by the same matrix, say P. Equation (127) is now,

$${}_a\mathrm{D}_x^{A} \, {}_a\mathrm{D}_x^{B} = \mathrm{P} \begin{pmatrix} {}_a\mathrm{D}_x^{\lambda_1} \, {}_a\mathrm{D}_x^{\rho_1} & & \\ & \mathrm{O} & \\ & & {}_a\mathrm{D}_x^{\lambda_m} \, {}_a\mathrm{D}_x^{\rho_m} \end{pmatrix} \mathrm{P}^{-1}. \tag{136}$$

A further simplification of equation (136) will occur using equations (5), (7), (8), (9), or (10) as the signs of the eigenvalues dictate. For example, if the eigenvalues of A and B are such that $\mathrm{Re}(\lambda_i) \le 0$ and $\mathrm{Re}(\rho_i) \le 0$ then equation (136) becomes,

$$ {}_a\mathrm{D}_x^{\mathrm{A}}\, {}_a\mathrm{D}_x^{\mathrm{B}} = \mathrm{P} \begin{pmatrix} {}_a\mathrm{D}_x^{\lambda_1+\rho_1} & & \\ & \circ & \\ & & {}_a\mathrm{D}_x^{\lambda_m+\rho_m} \end{pmatrix} \mathrm{P}^{-1} = {}_a\mathrm{D}_x^{\mathrm{A+B}}. \tag{137} $$

If the eigenvalues of matrices A and B are such that $\mathrm{Re}(\lambda_i) \le 0$ and $\mathrm{Re}(\rho_i) \le 0$ but not commuting then equation (127) is,

$$ {}_a\mathrm{D}_x^{\mathrm{A}}\, {}_a\mathrm{D}_x^{\mathrm{B}} = \mathrm{P} \left[ \mathrm{R}_{\mathrm{ij}}\; {}_a\mathrm{D}_x^{\lambda_\mathrm{i}+\rho_\mathrm{j}} \right] \mathrm{Q}^{-1}. \tag{138} $$

Or, using the spectral theorem,

$$ {}_a\mathrm{D}_x^{\mathrm{A}}\, {}_a\mathrm{D}_x^{\mathrm{B}} = \sum_{i=1}^{k} \sum_{j=1}^{l} G_i H_j\, {}_a\mathrm{D}_x^{\lambda_i+\rho_j}. \tag{139} $$

Let A, B $\in$ $\mathrm{N}_m$ with A = $\mathrm{PDP}^{\mathrm{T}}$ and B = $\mathrm{QEQ}^{\mathrm{T}}$ where D and E are diagonal. Denote the eigenvalues of A by $\lambda_i$ and the eigenvalues of B by $\rho_i$. Now consider the transpose of equation (127).

$$ \left( {}_a\mathrm{D}_x^{\mathrm{A}}\, {}_a\mathrm{D}_x^{\mathrm{B}} \right)^{\mathrm{T}} = \mathrm{Q} \begin{pmatrix} {}_a\mathrm{D}_x^{\rho_1} & & 0 \\ & \circ & \\ 0 & & {}_a\mathrm{D}_x^{\rho_m} \end{pmatrix} \mathrm{Q}^{\mathrm{T}}\, \mathrm{P} \begin{pmatrix} {}_a\mathrm{D}_x^{\lambda_1} & & 0 \\ & \circ & \\ 0 & & {}_a\mathrm{D}_x^{\lambda_m} \end{pmatrix} \mathrm{P}^{\mathrm{T}} \tag{140} $$

$$= {}_a D_x^{B} \, {}_a D_x^{A} \qquad (141)$$

As a final result consider the determinant of equation (119).

$$\det({}_a D_x^{A}) \; = \; \det(P) \left( \prod_{i=1}^{m} {}_a D_x^{\lambda_i} \right) \det(P^{-1}) \; = \; \prod_{i=1}^{m} {}_a D_x^{\lambda_i} \qquad (142)$$

The right hand side of equation (142) is a sequential fractional derivative (see e.g. [16] and [17]). Thus a sequential factional derivative may be viewed as the determinant of a matrix order fractional derivative. Additionally if the eigenvalues of A are such that $\mathrm{Re}(\lambda_i) \le 0$ then equation (142) can be simplified to give,

$$\det({}_a D_x^{A}) \; = {}_a D_x^{\mathrm{Tr(A)}} \qquad (143)$$

where Tr(A) is the trace of the matrix A, i.e. the sum of the eigenvalues.